\documentclass[aps,prl,reprint,amsmath,amssymb,nofootinbib,superscriptaddress]{revtex4-2}
\usepackage[T1]{fontenc}
\usepackage{microtype,mathrsfs}
\usepackage{empheq,tikz}
\usepackage{hyperref}
\newcommand{\F}{\mathcal F}

\newcommand{\bh}{\boldsymbol h}
\newcommand{\dd}{\mathrm d}

\newcommand{\cL}{c_{\mathrm L}}

\newcommand{\ns}{n_{\mathrm s}}

\hypersetup{
  colorlinks=true,
  linkcolor=blue,
  citecolor=blue,
  urlcolor=blue
}
\begin{document}
\title{Correlation functions of 2D critical models via a KPZ relation for conformal blocks}
\author{Gefei Cai}
\email{caigefei1917@pku.edu.cn}
\affiliation{Peking University, PR China}
\author{Shengzhi Jin}
\email{jinshengzhi@stu.pku.edu.cn}
\affiliation{Peking University, PR China}
\author{Xin Sun}
\email{xinsun@bicmr.pku.edu.cn}
\affiliation{Peking University, PR China}
\author{Baojun Wu}
\email{wubaojun@bit.edu.cn}
\affiliation{Beijing Institute of Technology, PR China}
\begin{abstract}
The Knizhnik--Polyakov--Zamolodchikov (KPZ) relation allows scaling exponents of two-dimensional (2D) critical systems to be derived from their counterparts in Liouville quantum gravity. We extend this framework to correlation functions based on an integral inversion relation between Virasoro conformal blocks with complementary central charges $c$ and $26-c$.
We demonstrate this method in various models by deriving the 2-point disk correlation from their quantum counterpart and the KPZ relation.
Results on critical loop models were previously known, but our method is robust enough to include many scenarios previously unsolved, such as the Fuzzy Potts model. 
Our approach is complementary to the conformal bootstrap.
\end{abstract}
\maketitle

\textit{Introduction.---}
A wide class of two-dimensional critical systems give rise to conformally invariant random fractals in the continuum limit. Important examples include percolation, loop model, polymer model and Brownian motion~\cite{Duplantier1989Fractals}. Since the 1980s, methods from conformal field theory (CFT) have been used to derive exact scaling dimensions and partition functions for these models \cite{Nienhuis-O-n,diFrancesco:1987qf}. For the critical loop model, a variety of correlation functions were recently obtained~\cite{NRJ24,DJNRS26,ACJNRSW26}. 
However, the nature of the underlying CFT usually has complicated features such as non-unitarity and non-rationality. Moreover, certain natural geometric observables do not fit into known CFT frameworks, such as the backbone exponent for 2D percolation~\cite{NQSZ-backbone-prl}. 
Coupling with 2D quantum gravity provides a complementary approach to these questions. The philosophy is that observables in the Euclidean geometry are related to their counterparts in the random geometry induced by quantum gravity, where the latter is usually more amenable to exact computation. At the level of scaling exponents, such relation is known as the Knizhnik--Polyakov--Zamolodchikov (KPZ) relation \cite{kpz-scaling}. Duplantier used this relation to derive the fractal dimension $4/3$ of the planar Brownian frontier \cite{duplantier-bm-exponents}. More recently, the coupling with 2D quantum gravity in the probabilistic framework \cite{DMS-mating-of-trees} together with the exact solvability of the Liouville CFT \cite{do-dozz,zz-dozz,krv-dozz} is used to derive the backbone exponent of 2D percolation, as well as certain annulus partition functions and bulk three-point structure constants for the critical loop model \cite{NQSZ-backbone-prl,ars-annuli,CFSX25,ACSW24ImDOZZ}. 

In this Letter, we further develop the quantum gravity approach by establishing a universal relation~\eqref{eq:kpz-reconstruction} between the disk two-point functions in the Euclidean and quantum setting. The crux of \eqref{eq:kpz-reconstruction} is a relation between Virasoro conformal blocks with complementary central charges $c$ and $26-c$, where conformal blocks play the role of wave functions in Fourier transform and its inversion. This relation is a consequence of the BRST cohomology for Virasoro algebra \cite{BMP92-BRST}. 

After presenting our main result~\eqref{eq:kpz-reconstruction} based on the block transform \eqref{eq:block-transformation}, we will give a derivation of \eqref{eq:block-transformation}. Then, we will present a variety of examples where \eqref{eq:kpz-reconstruction} is applied to derive the disk 2-point functions. This includes the diagonal operators and the two-leg operator for the critical loop model, and the two-point connectivity function for the Fuzzy Potts model. To our best knowledge, the case of Fuzzy Potts model was not solved before.

\textit{Main results.---}
Our main result reconstructs Euclidean correlation functions from their counterparts in quantum gravity. This reconstruction follows from an integral relation between Virasoro conformal blocks with complementary central charges and external weights.

Let $\bh=(h_1,h_2,h_3,h_4)$ be the external weights at $(0,x,1,\infty)$ with $0<x<1$, and set $\bh^\vee=(1-h_1,\ldots,1-h_4)$. We write the s-channel Virasoro conformal blocks as
\begin{equation}\label{eq:blocks}
\begin{aligned}
    \F_s^c(\bh;x) &\sim x^{s-h_1-h_2},\\
    \F_{1-s}^{26-c}(\bh^\vee;x) &\sim x^{h_1+h_2-s-1},
\end{aligned}
\end{equation}
where $s$ and $1-s$ are the respective internal weights; see the Supplementary Material (SM) for an explicit expression.
For a suitable class of test functions, we have the following block transform
\begin{equation}\label{eq:block-transformation}
\boxed{
\begin{split}
    \widehat f(s)&=\int_0^1 f(y) \, \F_{1-s}^{26-c}(\bh^\vee;y) \, \dd y, \\
    f(x)&=\frac{1}{2\pi i}\int_{\mathcal{C}} \widehat f(s) \, \F_s^c(\bh;x) \, \dd s.
\end{split}
}
\end{equation}
The integral contour $\mathcal C$ should be chosen suitably to make the integral converge. The convergence of the integral also depends on the asymptotic behavior of the test function $f$. For the applications below, where $c<1$, one valid choice is the parabolic contour
\begin{equation}
    \mathcal C_\delta: s(P) = (P+i\delta)^2 - \frac{1-c}{24}, \qquad P\in\mathbb R,
\end{equation}
oriented in the direction of increasing $P$, for $\delta>0$. This contour  was also used in the timelike Liouville theory \cite{Ribault-Santachiara}. 
The choice of the contour ensuring the convergence of the integral was discussed in \cite{JSW}.

Consider a critical system of central charge $c$ coupled to Liouville quantum gravity (LQG). We parametrize the Liouville theory by the coupling constant $b$, with background charge $Q=b+b^{-1}$ and central charge $\cL=26-c=1+6Q^2$. 
Let $\mathcal Z(x)$ be the disk two-point function with insertions of conformal weights $h_1,h_2$ at
\[ z_1=0, \qquad z_2=\sqrt{x}, \qquad 0<x<1. \]

We also consider its counterpart on Liouville quantum gravity.
We fix the bulk cosmological constant $\mu>0$, and use $\sigma$ to parametrize the boundary cosmological constant $\mu_B(\sigma)$ (see SM for the explicit parameterization).
The corresponding LQG surface can be described by a Liouville field $\Phi$, with quantum area measure $e^{2b\Phi(z)}\dd^2 z$. The Liouville insertions $V_{\alpha_j}(z_j)=e^{\alpha_j\Phi(z_j)}$ have conformal weights
\[ \Delta_{\alpha_j}=\frac{\alpha_j}{2}(Q-\frac{\alpha_j}{2})=1-h_j. \]
Let $\mathcal Q(\sigma)$ be the quantum two-point function, which is the coupled partition function integrated over the conformal modulus
\begin{equation}\label{eq:kpz-coupling}
    \mathcal Q(\sigma) = \int_0^1 \mathcal Z(x)\, \mathcal L_{\alpha_1,\alpha_2}(x;\sigma)\, \mathrm dx,
\end{equation}
where $\mathcal L_{\alpha_1,\alpha_2}(x;\sigma) = \left\langle V_{\alpha_1}(0)V_{\alpha_2}(\sqrt{x}) \right\rangle^{\mathrm{LF}}_{\mathbb D,\mu,\mu_B(\sigma)}$ is the Liouville two-point function.

The Liouville correlation is determined by the bootstrap expansion \cite{do-dozz,zz-dozz,FZZ},
\begin{equation}\label{eq:liouville-bootstrap}
    \mathcal L_{\alpha_1,\alpha_2}(x;\sigma)
    = \int_0^\infty \cos(\pi\sigma p)\, C_{\alpha_1,\alpha_2}(p)\, \F_{\Delta_{Q+ip}}^{26-c}(\bh^\vee;x) \, \mathrm dp.
\end{equation}
The coefficient $C_{\alpha_1,\alpha_2}(p)$ combines the DOZZ three-point structure constant and the $\sigma$-independent part of the FZZ one-point structure constant
\begin{equation}
    C_{\alpha_1,\alpha_2}(p) = \frac{1}{4\pi} C_{\mathrm{DOZZ}}(\alpha_1,\alpha_2,Q+ip) \Psi_{\mathrm{FZZ}}(-p).
\end{equation}
See SM for expressions of the DOZZ and FZZ formula.

Denote the inverse Fourier transform of $\mathcal{Q}$ by
\begin{equation}\label{eq:inverse-cosine-transform}
    \widetilde{\mathcal Q}(p) = 2\int_0^\infty \mathcal Q(\sigma) \cos(\pi\sigma p) \,\dd\sigma.
\end{equation}
Substituting \eqref{eq:liouville-bootstrap} into \eqref{eq:kpz-coupling} and using the block transform \eqref{eq:block-transformation}, we have
\begin{equation}\label{eq:kpz-reconstruction}
\boxed{
    \mathcal Z(x) = \frac{1}{2\pi i} \int_{\mathcal{C}} \frac{\widetilde{\mathcal Q}(p(s))} {C_{\alpha_1,\alpha_2}(p(s))} \F_s^c(\bh;x)\,\mathrm ds.
}
\end{equation}
Here $s+\Delta_{Q+ip(s)}=1$ hence $p(s) = \sqrt{4(1-s)-Q^2}$,
and the external weights for the blocks are $\bh=(h_1,h_2,h_2,h_1)$ and $\bh^\vee=(1-h_1,1-h_2,1-h_2,1-h_1)$.

Equation~\eqref{eq:kpz-reconstruction} recovers the Euclidean correlation function from its quantum counterpart. It is a bulk-channel expansion on conformal blocks.
The integral kernel in \eqref{eq:kpz-reconstruction} is universal. Thus, this relation applies to any critical system whose quantum two-point correlation function is computable.

\textit{Derivation of the block transform \eqref{eq:block-transformation}.---}
We now prove an algebraic identity underlying \eqref{eq:block-transformation} for four independent external weights. Write
\begin{equation}\label{eq:block-coefficient}
\begin{aligned}
    \F_s^c(\bh;x) {}&= x^{s-h_1-h_2} \sum_{n=0}^{\infty} B_n^c(s;\bh)x^n, \\
    \F_{1-s}^{26-c}(\bh^\vee;x) {}&= x^{h_1+h_2-s-1} \sum_{m=0}^{\infty} B_m^{26-c}(1-s;\bh^\vee)x^m;
\end{aligned}
\end{equation}
see SM for the explicit expression.
For integer $n\ge0$ and internal weights satisfying $s_1+s_2=1-n$, we now argue
\begin{equation}\label{eq:finite-level-identity}
    \sum_{j=0}^n B_j^c(s-n;\bh) B_{n-j}^{26-c}(1-s;\bh^\vee)=\delta_{n0}.
\end{equation}

The case $n=0$ is directly from $B_0^c=B^{26-c}_0=1$.
For $n\ge1$, consider the tensor product of Verma modules $V_{c,s_1}\otimes V_{26-c,s_2}$.
Let $v_j^{h^*_1,h^*_2}(c^*,s^*)$ denote the level-$j$ OPE vector for external weights $(h^*_1,h^*_2)$ in the internal Verma module $V_{c^*,s^*}$, normalized by $v_0^{h^*_1,h^*_2}(c^*,s^*)=|s^*\rangle$. 
We denote by $W_n^{h_1^*,h_2^*}$ the component of total level $n$ in the tensor product of the two OPE vectors:
\begin{equation}\label{eq:OPE-tensor}
    W_n^{h_1^*,h_2^*}=\sum_{j=0}^n v_j^{h_1^*,h_2^*}(c,s_1) \otimes v_{n-j}^{1-h_1^*,1-h_2^*}(26-c,s_2).
\end{equation}
By the representation of block coefficients as Shapovalov pairings, the sum in Equation~\eqref{eq:finite-level-identity} equals $\langle W_n^{h_4,h_3},W_n^{h_1,h_2}\rangle$, where the brackets denote the product Shapovalov form.

Define the diagonal actions by
\[ L_m^{\mathrm{diag}}=L_m\otimes1+1\otimes L_m. \]
They form a Virasoro algebra of central charge 26, and its total-level-$n$ vectors have weight $s_1+s_2+n=1$. 
The OPE vectors satisfy the Ward recursion for $m>0$
\[ L_m v_j^{h_1^*,h_2^*}(c^*,s^*) = (s^*+j-m+mh_2^*-h_1^*)\,v_{j-m}^{h_1^*,h_2^*}(c^*,s^*). \]
The complementary external weights thus yield
\begin{equation}\label{eq:primary-vector}
    L_0^{\mathrm{diag}} W_n^{h_1^*,h_2^*} = W_n^{h_1^*,h_2^*}, \qquad L_m^{\mathrm{diag}} W_n^{h_1^*,h_2^*} = 0, \quad m\ge1,
\end{equation}
i.e. $W_n^{h_1^*,h_2^*}$ is a primary vector of weight 1 in the tensor module.

We now couple the tensor module to a $bc$ ghost system of central charge $-26$. We consider the tensor product of $W_n^{h_1,h_2}$ with the ghost vacuum. By \eqref{eq:primary-vector}, this is a BRST-closed state. The relative BRST cohomology vanishes at all positive levels, hence this state is also exact \cite{BMP92-BRST}. 
By extracting the ghost-vacuum component of the exactness relation, we have that $W_n^{h_1,h_2}$ is a sum of descendants in the tensor module
\[ W_n^{h_1,h_2} = \sum_{m=1}^{n} L_{-m}^{\mathrm{diag}}U_m^{h_1,h_2} \]
with some $U_m^{h_1,h_2}$ in the tensor module.
Using the adjoint relation for the Shapovalov form and \eqref{eq:primary-vector}, we obtain
\[ \langle W_n^{h_4,h_3},W_n^{h_1,h_2}\rangle = \sum_{m=1}^n \langle L_m^{\mathrm{diag}} W_n^{h_4,h_3}, U_m^{h_1,h_2} \rangle = 0, \]
which proves \eqref{eq:finite-level-identity} for $n\ge1$.

We now derive the integral transform~\eqref{eq:block-transformation} from \eqref{eq:finite-level-identity}. Set $s_1=s-n$ and $s_2=1-s$.
Consider the infinite upper-triangular matrices $U(u)$ and $V(u)$ defined by 
\[
\begin{aligned}
    U_{rk}(u)&=B_{k-r}^c(u-k;\bh),\\
    V_{rk}(u)&=B_{k-r}^{26-c}(1-u+r;\bh^\vee),
\end{aligned}
\]
for $0\le r\le k$, with all other entries zero. The finite-level identity \eqref{eq:finite-level-identity} gives $(VU)_{rk}=\delta_{rk}$. Since both matrices are upper-triangular, this yields $VU=I=UV$.
We then expand $\F_s^c(\bh;x) \F_{1-s}^{26-c}(\bh^\vee;y)$ by \eqref{eq:block-coefficient}, and arrange the terms by the total level $N$. Set $u=s+n$ where $n$ denotes the level in the expansion of $\F_s^c(\bh;x)$. The contribution of any level $N>0$ contains a factor $(UV)_{0N}$, and thus vanishes. The remaining term at level $N=0$ is the Mellin inversion kernel:
\begin{equation}
\begin{aligned}
    & \frac{1}{2\pi i}\int \F_s^c(\bh;x) \, \F_{1-s}^{26-c}(\bh^\vee;y) \, \dd s \\
    = {}& \sum_{N=0}^{\infty} \frac{y^{N-1}}{2\pi i} \int \left(\frac{x}{y}\right)^{u-h_1-h_2}\,(UV)_{0N}(u) \,\dd u 
    = \delta(x-y).
\end{aligned}
\end{equation}
This proves the block transform \eqref{eq:block-transformation}.

\textit{Examples.---}
We now demonstrate how Equation~\eqref{eq:kpz-reconstruction} can be used to derive two-point functions in critical models by several examples.

\begin{figure}[!h]
    \centering
    \includegraphics[width=0.70\columnwidth]{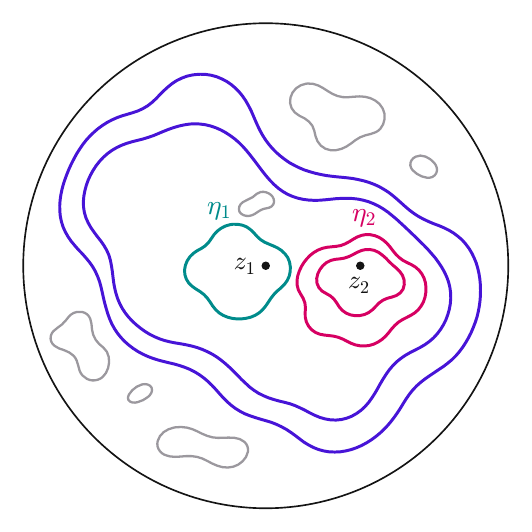}
    \caption{Loop nesting in the $O(n)$ model. Green, red, purple, and gray loops carry weights $n_1$, $n_2$, $n_{\mathrm{s}}$, and $n$, respectively, according to their nesting type. The outermost green and red loops are denoted by $\eta_1$ and $\eta_2$, respectively.}
    \label{fig:O(n)-nesting}
\end{figure}

\textit{Diagonal operators in the loop model.}
Consider the dilute $O(n)$ loop model in the disk with two marked points $0$ and $\sqrt{x}$. We assign fugacities $n_1,n_2,n_{\mathrm s}$ to loops surrounding only the first marked site, only the second, or both, respectively, while all other loops retain fugacity $n$; see Figure~\ref{fig:O(n)-nesting} for an illustration. Its continuum limit can be described by the nested conformal loop ensemble (CLE$_\kappa$) with parameter
\[ n=-2\cos(\pi\beta^2), \qquad \kappa=\frac{4}{\beta^2}, \qquad c=1-6(\beta-\beta^{-1})^2. \]

After local renormalization, the local nesting weights are encoded by the conformal radius \cite{ACSW24ImDOZZ}.
Let $\eta_j$ be the outermost loop in CLE$_\kappa$ surrounding only the $j$-th mark, and let $f_j$ be a conformal map from the unit disk to the domain bounded by $\eta_j$ with $f_j(0)=z_j$. The conformal radius of $\eta_j$ viewed from $z_j$ is defined as $R_j=|f_j'(0)|$. Let $N$ count the loops surrounding both points. The continuum two-point function is then defined by 
\begin{equation}\label{eq:CLE-correlation}
    \mathcal{Z}^{O(n)}_{h_1,h_2,\ns}(x) = \mathbb{E}_{\mathrm{CLE}_\kappa}\left[ R_1^{-2h_1} R_2^{-2h_2} \left(\frac{n_s}{n}\right)^N \right],
\end{equation}
where the local fugacities and conformal weights are parametrized by momenta $P_j$ through
\begin{equation}\label{eq:CLE-dimension}
    n_j = 2\cos(2\pi\beta P_j), \qquad h_j = P_j^2-\frac{1}{4}(\beta-\beta^{-1})^2.
\end{equation}

\begin{figure}[!h]
    \centering
    \includegraphics[width=\columnwidth]{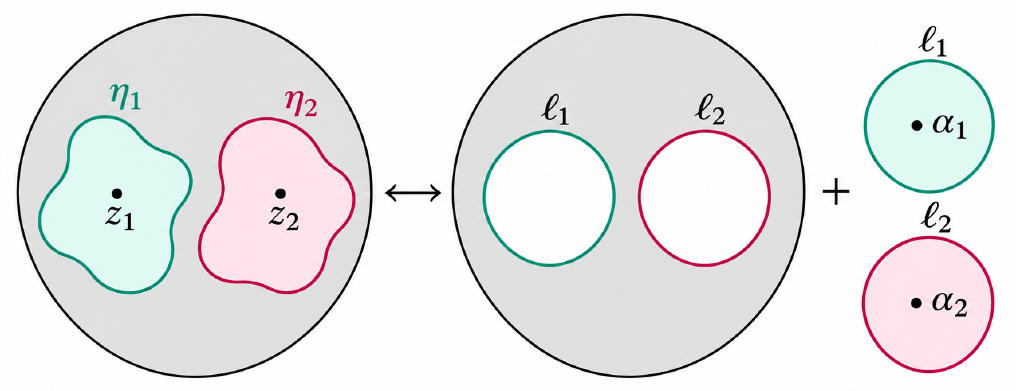}
    \caption{Cutting along $\eta_1$ and $\eta_2$ decomposes the quantum disk into a quantum pair-of-pants and two quantum disks with bulk charges $\alpha_1$ and $\alpha_2$, respectively.}
    \label{fig:cutting}
\end{figure}

We present the computation for its quantum counterpart in the dilute regime $\beta\in[1,\sqrt{3/2})$, $\kappa\in(8/3,4]$. Consider the two-pointed Liouville quantum disk with coupling constant $b=\beta^{-1}$ and charges 
\[ \alpha_j=Q-2P_j \]
satisfying $h_j+\Delta_{\alpha_j}=1$. We independently couple a CLE with the two-pointed quantum disk. Cutting the quantum disk along $\eta_1,\eta_2$ defined above separates the surface into three parts, as shown in Figure~\ref{fig:cutting}: two quantum disks, each with one bulk marked point with charge $\alpha_j$, and a quantum pair-of-pants. Conditional on the quantum boundary lengths of $\eta_1$ and $\eta_2$, these three surfaces are independent \cite{msw-cle-lqg,ACSW24ImDOZZ}. 
Thus, we have
\begin{equation}
    \mathcal{Q}^{O(n)}_{\alpha_1, \alpha_2, n_{\mathrm s}}(\sigma)
    =\! \int_0^\infty\!\! \int_0^\infty \!P^{n_{\mathrm s}}(\ell_1,\ell_2;\sigma) \! \prod_{j=1,2} \! U(\alpha_j,\ell_j)\ell_j\mathrm d\ell_j.
\end{equation}
Here, $U(\alpha;\ell)$ denotes the fix-length FZZ formula \cite{FZZ} (see SM). $P^{n_{\mathrm s}}(\ell_1,\ell_2;\sigma)$ denotes the partition function of the quantum pair-of-pants whose two inner boundary lengths are fixed at $\ell_1,\ell_2$ and whose outer boundary is weighted by the boundary cosmological constant $\mu_B(\sigma)$. Its formula can be computed exactly \cite{ACSW24ImDOZZ}. The factor $\ell_j$ is contributed from the uniform gluing along the boundary.
Combining these gives
\[ 
    \mathcal{Q}^{O(n)}_{\alpha_1, \alpha_2, n_{\mathrm s}}(\sigma) = \mathcal K_{\alpha_1,\alpha_2} \int_0^\infty \frac{\cos(\pi\sigma p)} {\cosh(\pi\beta p)-n_{\mathrm s}/2}\,\mathrm dp,
\]
where $K_{\alpha_1,\alpha_2}$ is some explicit constant independent of $p$ and $\sigma$.

Applying Equation~\eqref{eq:kpz-reconstruction} then yields
\begin{equation}\label{eq:O(n)-nesting-formula}
\begin{aligned}
    \mathcal Z^{O(n)}_{h_1,h_2,\ns}(x)
    = \frac{\mathcal{N}_{P_1,P_2}}{2\pi i} \int_{\mathcal C} \frac{\sin(2\pi\beta^{-1}P(s)) \, \sin(2\pi\beta P(s))}{P(s) \left[\cos(2\pi\beta P(s)) - n_{\mathrm{s}}/2\right]} \\
    \times \, \hat{C}^{P(s)}_{(0,2\beta P_1)(0,2\beta P_2)} \, \mathcal{F}_{s}^c(h_1,h_2,h_2,h_1;x) \, \dd s.
\end{aligned}
\end{equation}
Here and in \eqref{eq:O(n)-loop-connectivity} and \eqref{eq:Fuzzy-Potts-connectivity}, $P(s)=\sqrt{s+\frac{1}{4}(\beta-\beta^{-1})^2}$, and $\hat{C}$ is the three-point structure constant in the imaginary-DOZZ normalization. The normalizing factor $\mathcal N_{P_1,P_2}$ is independent of $x,s$. See SM for their formulas.

This integral formula \eqref{eq:O(n)-nesting-formula} agrees with the result in \cite{JSW} from the SLE-loop CFT derivation. Evaluating the integral by residues recovers the bootstrap expansion with discrete spectrum as in~\cite{DJNRS26}; see SM for the explicit formula.
Putting $n_1=n_2=0$ and $n_\mathrm{s}=n,-n$ gives the connectivity probability for $O(n)$ and FK random cluster models. 

Equation~\eqref{eq:O(n)-nesting-formula} also extends to the dense regime $\beta\in(1/\sqrt{2},1)$, $\kappa\in(4,8)$. In this regime, CLE loops can touch themselves and one another, and will cut the surface into generalized quantum surfaces. Despite the more complicated geometry, the same strategy applies to the computation of the quantum two-point function and yields the same expression~\eqref{eq:O(n)-nesting-formula}.

\textit{Two-leg operator in the loop model.}
Another example is the probability that two points are on the same $O(n)$ loop, which corresponds to the two-leg operator for the loop model. Each insertion has conformal weight $h_{(1,0)}=\frac{1}{2}-\frac{1}{4\beta^2}$. 
Its continuum limit can be described by a $(2-2h_{(1,0)})$-dimensional measure on CLE loops \cite{ACSW24ImDOZZ}.

Cutting the quantum disk along the loop passing through the two points, we get a quantum annulus and a quantum disk with two boundary marked points.  We can therefore compute the quantum correlation function and by \eqref{eq:kpz-reconstruction} yield
\begin{equation}\label{eq:O(n)-loop-connectivity}
\begin{aligned}
    \mathcal{Z}^{O(n)}_{(1,0)(1,0)}(x) = \frac{\mathcal N}{2\pi i} \int_{\mathcal{C}} \frac{\sin(2\pi\beta^{-1}P(s)) \, \sin(2\pi\beta P(s))}{P(s) \left[\cos(2\pi\beta P(s)) - n/2\right]} \\
    \times \, \hat{C}^{P(s)}_{(1,0)(1,0)} \, \mathcal F_s^c(h_{(1,0)}, h_{(1,0)}, h_{(1,0)}, h_{(1,0)}; x) \, \dd s.
\end{aligned}
\end{equation}
Here and in \eqref{eq:Fuzzy-Potts-connectivity}, $\mathcal{N}$ is a constant arising from the choice of normalization for the continuum measure on the corresponding fractals, and is independent of $s$ and $x$.
Evaluating by residues also recovers the discrete summation which agrees with \cite[Formula (2.8)]{DJNRS26}.

\textit{Spin-cluster connectivity in the Fuzzy Potts model.}
Consider the critical FK random-cluster model in $\mathbb{D}$ with  
\[ q=4\cos^2(\pi \beta^{-2})\in(1,4), \qquad \beta\in(1,\sqrt{3/2}). \]
We independently color each FK cluster red with probability $1/q$ and blue with probability $1-1/q$, and merge adjacent clusters of the same color. For integer $q$, the red clusters described the sites of a $q$-Potts model with a distinguished spin value.
We consider the all-red and all-blue boundary conditions, denoted by $\mathrm{R}$ and $\mathrm{B}$ respectively.
In the continuum limit, the interfaces between spin clusters can be described by CLE and its variants \cite{cle-percolations}.

We consider the probability that two points are in the same red cluster, whose fractal dimension is $(2-2h_{(0,1/2)})$ with $h_{(0,1/2)}=\frac{1}{2}-\frac{\beta^2}{4}-\frac{3}{16\beta^2}$.
For the two boundary conditions $\mathrm{bc} \in \{\mathrm{R},\mathrm{B}\}$, we can compute the quantum 2-point function following the method in~\cite{CLWZ25}. 
Applying~\eqref{eq:kpz-reconstruction}, we obtain
\begin{equation}\label{eq:Fuzzy-Potts-connectivity}
\begin{aligned}
    \mathcal{Z}^{\mathrm{bc}}(x) = \frac{\mathcal N}{2\pi i} \int_{\mathcal{C}} \frac{\bigl[\sin(4u-j\theta)-2\chi\sin(2u-j\theta)\bigr]}{\bigl[-2\cos(\pi\beta^2)\bigr]^{j-1}  \,\bigl[\cos(2u)-\chi\bigr]} \\
    \times \, \frac{\sin\theta}{P(s)\cos u} \, \hat{C}^{P(s)}_{(0,\frac{1}{2})(0,\frac{1}{2})} \, \mathcal F_s^c(h_{(0,\frac{1}{2})}, \cdots , h_{(0,\frac{1}{2})}; x) \, \dd s,
\end{aligned}
\end{equation}
where $u=2\pi\beta^{-1}P(s)$, $\theta=2\pi\beta P(s)$, $\chi=q/2-1=\cos(2\pi\beta^{-2})$, $j=1$ for $\mathrm{bc}=\mathrm{R}$ and $j=2$ for $\mathrm{bc}=\mathrm{B}$.

\textit{Conclusion.---}
To summarize, we establish a universal relation that reconstructs disk two-point functions of critical systems from their quantum-gravity counterpart. This relation is based on the integrability of Liouville CFT and an integral transform \eqref{eq:kpz-reconstruction} involving conformal blocks with complementary central charges and conformal weights.
We use it to compute two-point correlation functions in the $O(n)$ loop model with both diagonal and non-diagonal insertions, as well as the spin-cluster connectivity in the Fuzzy Potts model. This method is quite general, valid as long as the quantum correlation function is computable.
In particular, this method can be applied to the two-point connectivity of the backbone event in critical percolation, and to loop models with more general boundary conditions.

The derivation does not presume the operator product expansion of the matter fields. The resulting formula is an integral of conformal blocks along an auxiliary spectral contour. For certain observables such as \eqref{eq:O(n)-nesting-formula} and \eqref{eq:O(n)-loop-connectivity}, evaluating the integral by residues gives the discrete bootstrap expansion, and recovers the structure constants and discrete spectrum which are predicted from the known CFT structure. 
For other observables such as the Fuzzy Potts connectivity \eqref{eq:Fuzzy-Potts-connectivity} and the percolation backbone connectivity, the CFT interpretation of the integral remains open. 

Finally, this paper is one step of our program to study the CFT underlying the critical loop model through quantum gravity. We will report our progress on the four-point sphere correlation in a forthcoming work.

\smallskip
\noindent{\bf Acknowledgments.}
We thank Jesper Lykke Jacobsen and Rongvoram Nivesvivat for helpful discussions. 
G.C., S.J., X.S. and B.W.\ are  supported by National Key R\&D Program of China (No.\ 2023YFA1010700) and the National Natural Science Foundation of China Grant (No. 12526204).

\noindent{\bf AI Disclosure.}
Building on~\cite{ACSW24ImDOZZ,ars-annuli}, we developed the conceptual framework of this paper. While working on the closely related project \cite{JSW}, we identified the key formula~\eqref{eq:block-transformation}. The derivation of~\eqref{eq:block-transformation} based on~\cite{BMP92-BRST} was developed by AI. We take full responsibility for the entire content of the paper.
\bibliography{cibib}

@article{Duplantier1989Fractals,
  author  = {Duplantier, Bertrand},
  title   = {Fractals in two dimensions and conformal invariance},
  journal = {Physica D},
  volume  = {38},
  pages   = {71--87},
  year    = {1989},
  doi     = {10.1016/0167-2789(89)90175-9}
}

@article{Nienhuis-O-n,
  author  = {Nienhuis, Bernard},
  title   = {Exact critical point and critical exponents of {$O(n)$} models in two dimensions},
  journal = {Phys. Rev. Lett.},
  volume  = {49},
  pages   = {1062--1065},
  year    = {1982},
  doi     = {10.1103/PhysRevLett.49.1062}
}

@article{diFrancesco:1987qf,
  author  = {di Francesco, Philippe and Saleur, Hubert and Zuber, Jean-Bernard},
  title   = {Relations between the {Coulomb} gas picture and conformal invariance of two-dimensional critical models},
  journal = {J. Stat. Phys.},
  volume  = {49},
  pages   = {57--79},
  year    = {1987},
  doi     = {10.1007/BF01009954}
}

@article{NRJ24,
  author  = {Nivesvivat, Rongvoram and Ribault, Sylvain and Jacobsen, Jesper Lykke},
  title   = {Critical loop models are exactly solvable},
  journal = {SciPost Phys.},
  volume  = {17},
  pages   = {029},
  year    = {2024},
  doi     = {10.21468/SciPostPhys.17.2.029}
}

@misc{DJNRS26,
  author        = {Downing, Max and Jacobsen, Jesper Lykke and Nivesvivat, Rongvoram and Ribault, Sylvain and Saleur, Hubert},
  title         = {Reflecting boundary conditions in critical loop models},
  year          = {2025},
  eprint        = {2512.10400},
  archivePrefix = {arXiv},
  primaryClass  = {hep-th}
}

@misc{ACJNRSW26,
  author        = {Ang, Morris and Cai, Gefei and Jacobsen, Jesper Lykke and Nivesvivat, Rongvoram and Roux, Paul and Sun, Xin and Wu, Baojun},
  title         = {Exact solution of three-point functions in critical loop models},
  year          = {2026},
  eprint        = {2604.05503},
  archivePrefix = {arXiv},
  primaryClass  = {cond-mat.stat-mech}
}

@article{NQSZ-backbone-prl,
  author  = {Nolin, Pierre and Qian, Wei and Sun, Xin and Zhuang, Zijie},
  title   = {Backbone exponent and annulus crossing probability for planar percolation},
  journal = {Phys. Rev. Lett.},
  volume  = {134},
  pages   = {117101},
  year    = {2025},
  doi     = {10.1103/PhysRevLett.134.117101}
}

@article{kpz-scaling,
  author  = {Knizhnik, V. G. and Polyakov, A. M. and Zamolodchikov, A. B.},
  title   = {Fractal structure of {2D}-quantum gravity},
  journal = {Mod. Phys. Lett. A},
  volume  = {3},
  pages   = {819--826},
  year    = {1988},
  doi     = {10.1142/S0217732388000982}
}

@article{duplantier-bm-exponents,
  author  = {Duplantier, Bertrand},
  title   = {Random walks and quantum gravity in two dimensions},
  journal = {Phys. Rev. Lett.},
  volume  = {81},
  pages   = {5489--5492},
  year    = {1998},
  doi     = {10.1103/PhysRevLett.81.5489}
}

@book{DMS-mating-of-trees,
  author    = {Duplantier, Bertrand and Miller, Jason and Sheffield, Scott},
  title     = {{Liouville} Quantum Gravity as a Mating of Trees},
  series    = {Ast{\'e}risque},
  volume    = {427},
  publisher = {Soci{\'e}t{\'e} Math{\'e}matique de France},
  address   = {Paris},
  year      = {2021},
  doi       = {10.24033/ast.1149}
}

@article{do-dozz,
  author  = {Dorn, H. and Otto, H.-J.},
  title   = {Two- and three-point functions in {Liouville} theory},
  journal = {Nucl. Phys. B},
  volume  = {429},
  pages   = {375--388},
  year    = {1994},
  doi     = {10.1016/0550-3213(94)00352-1}
}

@article{zz-dozz,
  author  = {Zamolodchikov, A. B. and Zamolodchikov, {\relax Al}. B.},
  title   = {Conformal bootstrap in {Liouville} field theory},
  journal = {Nucl. Phys. B},
  volume  = {477},
  pages   = {577--605},
  year    = {1996},
  doi     = {10.1016/0550-3213(96)00351-3}
}

@article{krv-dozz,
  author  = {Kupiainen, Antti and Rhodes, R{\'e}mi and Vargas, Vincent},
  title   = {Integrability of {Liouville} theory: proof of the {DOZZ} formula},
  journal = {Ann. Math.},
  volume  = {191},
  pages   = {81--166},
  year    = {2020},
  doi     = {10.4007/annals.2020.191.1.2}
}

@article{ars-annuli,
  author  = {Ang, Morris and Remy, Guillaume and Sun, Xin},
  title   = {The moduli of annuli in random conformal geometry},
  journal = {Ann. Sci. \'{E}c. Norm. Sup\'{e}r.},
  volume  = {58},
  pages   = {1037--1087},
  year    = {2025},
  doi     = {10.24033/asens.2624}
}

@misc{CFSX25,
  author        = {Cai, Gefei and Fu, Xuesong and Sun, Xin and Xie, Zhuoyan},
  title         = {Disconnection probability of {Brownian} motion on an annulus},
  year          = {2025},
  eprint        = {2509.14073},
  archivePrefix = {arXiv},
  primaryClass  = {math.PR}
}

@article{ACSW24ImDOZZ,
  author  = {Ang, Morris and Cai, Gefei and Sun, Xin and Wu, Baojun},
  title   = {Integrability of conformal loop ensemble: imaginary {DOZZ} formula and beyond},
  journal = {J. Am. Math. Soc.},
  year    = {2026},
  doi     = {10.1090/jams/1086}
}

@article{BMP92-BRST,
  author  = {Bouwknegt, Peter and McCarthy, Jim and Pilch, Krzysztof},
  title   = {{BRST} analysis of physical states for {2D} gravity coupled to {$c \leq 1$} matter},
  journal = {Commun. Math. Phys.},
  volume  = {145},
  pages   = {541--560},
  year    = {1992},
  doi     = {10.1007/BF02099397}
}

@article{Ribault-Santachiara,
  author  = {Ribault, Sylvain and Santachiara, Raoul},
  title   = {{Liouville} theory with a central charge less than one},
  journal = jhep,
  volume  = {2015},
  number  = {08},
  pages   = {109},
  year    = {2015},
  doi     = {10.1007/JHEP08(2015)109}
}

@misc{JSW,
  author        = {Jin, Shengzhi and Sun, Xin and Wu, Baojun},
  title         = {Disk 2-point functions for conformal loop ensemble: a derivation from the {SLE} loop {CFT}},
  year          = {2026},
  eprint        = {2611.27767},
  archivePrefix = {arXiv},
  primaryClass  = {cond-mat.stat-mech}
}

@misc{FZZ,
  author        = {Fateev, V. and Zamolodchikov, A. and Zamolodchikov, {\relax Al}.},
  title         = {Boundary {Liouville} field theory {I}. {Boundary} state and boundary two-point function},
  year          = {2000},
  eprint        = {hep-th/0001012},
  archivePrefix = {arXiv}
}

@article{msw-cle-lqg,
  author  = {Miller, Jason and Sheffield, Scott and Werner, Wendelin},
  title   = {Simple conformal loop ensembles on {Liouville} quantum gravity},
  journal = {Ann. Probab.},
  volume  = {50},
  pages   = {905--949},
  year    = {2022},
  doi     = {10.1214/21-AOP1550}
}

@article{cle-percolations,
  author  = {Miller, Jason and Sheffield, Scott and Werner, Wendelin},
  title   = {{CLE} percolations},
  journal = {Forum Math. Pi},
  volume  = {5},
  pages   = {e4},
  year    = {2017},
  doi     = {10.1017/fmp.2017.5}
}

@misc{CLWZ25,
  author        = {Cai, Gefei and Liu, Haoyu and Wu, Baojun and Zhuang, Zijie},
  title         = {Three-point connectivity constant for {$q$}-state {Potts} spin clusters},
  year          = {2025},
  eprint        = {2510.05850},
  archivePrefix = {arXiv},
  primaryClass  = {math.PR}
}

\clearpage
\onecolumngrid
\newpage

\section{Supplementary Material}

\subsection{The functions $\Gamma_{\hat{b}}(z)$ and $\Upsilon_{\hat{b}}(z)$}

We first present the formula for the Barnes double Gamma function appearing in the expression of three-point structure constants. 
For $\mathrm{Re}(\hat{b})>0$, we introduce $\hat{Q}=\hat{b}+\hat{b}^{-1}$.

The Barnes double Gamma function is fixed by $\Gamma_{\hat{b}}(\hat{Q}/2)=1$ and the shift equations
\begin{equation}\label{sm:Barnes-shifts}
\begin{aligned}
    \Gamma_{\hat{b}}(z+\hat{b}) &= \frac{\sqrt{2\pi}\,\hat{b}^{\hat{b}z-1/2}}{\Gamma(\hat{b}z)}\Gamma_{\hat{b}}(z),\\
    \Gamma_{\hat{b}}(z+\hat{b}^{-1}) &= \frac{\sqrt{2\pi}\,\hat{b}^{-z/\hat{b}+1/2}}{\Gamma(z/\hat{b})}\Gamma_{\hat{b}}(z).
\end{aligned}
\end{equation}
Its expression is given by the double Gamma function
\begin{equation}
    \Gamma_{\hat{b}}(z) = \Gamma_{\hat{b}^{-1}}(z) = \frac{\Gamma_2(z|\hat{b},\hat{b}^{-1})}{\Gamma_2(\hat{Q}/2|\hat{b},\hat{b}^{-1})}
\end{equation}
Define
\begin{equation}\label{sm:upsilon}
    \Upsilon_{\hat{b}}(z)=\frac{1}{\Gamma_{\hat{b}}(z)\Gamma_{\hat{b}}(\hat{Q}-z)}.
\end{equation}
It satisfies the shift equations and the symmetry equation
\begin{equation}\label{sm:upsilon-shifts}
\begin{aligned}
    \Upsilon_{\hat{b}}(z+\hat{b}) &= \gamma(\hat{b}z)\hat{b}^{1-2\hat{b}z}\Upsilon_{\hat{b}}(z),\\
    \Upsilon_{\hat{b}}(z+\hat{b}^{-1}) &= \gamma(\hat{b}^{-1}z)\hat{b}^{2z/\hat{b}-1}\Upsilon_{\hat{b}}(z),\\
    \Upsilon_{\hat{b}}(z)&=\Upsilon_{\hat{b}}(\hat{Q}-z).
\end{aligned}
\end{equation}
where $\gamma(z)=\Gamma(z)/\Gamma(1-z)$ and $\Gamma$ is the Gamma function.

\subsection{The Liouville structure constants}

\paragraph{The FZZ formula.}

We disintegrate the disk one-point Liouville field measure over the quantum length, and write $U(\alpha;\ell)$ for the partition function of the fix-length measure, which is given by the FZZ formula
\begin{equation}\label{sm:FZZ-fixed-length}
    U(\alpha;\ell) = \frac{2}{b} \left[\pi\mu\gamma(b^2)\right]^{(Q-\alpha)/(2b)} \frac{\Gamma(1-(Q-\alpha)b)}{\Gamma((Q-\alpha)/b)} \frac{K_{(Q-\alpha)/b}(\sqrt{\mu/\sin(\pi b^2)}\ell)}{\ell},
\end{equation}
where $K_\nu$ is the modified Bessel function of the second kind
\begin{equation}
    K_\nu(x) = \int_0^\infty e^{-x\cosh t} \cosh(\nu t) \, \dd t.
\end{equation}

The FZZT boundary parameterization is defined by
\begin{equation}\label{sm:muB}
    \mu_B(\sigma) = \sqrt{\frac{\mu}{\sin(\pi b^2)}}\cosh(\pi b\sigma).
\end{equation}
The Liouville disk one-point function is 
\begin{equation}
\begin{aligned}
    U_\sigma(Q+ip) &= \left\langle V_{Q+ip}(0)\right\rangle^{\mathrm{LF}}_{\mathbb D,\mu,\mu_B(\sigma)} = \int_0^\infty e^{-\mu_B(\sigma)\ell} \, U(Q+ip;\ell) \, \dd\ell
    = \cos(\pi\sigma p) \Psi_{\mathrm{FZZ}}(p), \\
    \Psi_{\mathrm{FZZ}}(p) &= 2[\pi\mu\gamma(b^2)]^{-ip/(2b)} \frac{\Gamma(1+ib^{-1}p)\Gamma(1+ibp)}{ip}.
\end{aligned}
\end{equation}

\paragraph{The DOZZ formula.}
Let $\alpha=\alpha_1+\alpha_2+\alpha_3$. The Liouville sphere three-point function is 
\begin{equation}
\begin{aligned}
    & C_{\mathrm{DOZZ}}(\alpha_1,\alpha_2,\alpha_3) := \left\langle V_{\alpha_1}(0) V_{\alpha_2}(1) V_{\alpha_3}(\infty) \right\rangle^{\mathrm{LF}}_{\mathbb S^2,\mu} \\
    =& \left[\pi\mu\gamma(b^2)b^{2-2b^2}\right]^{(2Q-\alpha)/(2b)} \frac{\Upsilon_b'(0) \prod_{j=1}^3 \Upsilon_b(\alpha_j)} {\Upsilon_b(\alpha/2-Q) \prod_{j=1}^3 \Upsilon_b(\alpha/2-\alpha_j)}.
\end{aligned}
\end{equation}

\subsection{The imaginary DOZZ formula}

We use the imaginary DOZZ formula in the following normalization
\begin{equation}
    \hat C_{(r_1, s_1)(r_2, s_2)}^P =
    \frac{16P \sin(\pi\beta^2) \sin(\pi\beta^{-2}) \Gamma_\beta(2P) \Gamma_\beta(-2P)}{\displaystyle\prod_{\epsilon_1,\epsilon_2,\epsilon_3=\pm} \Gamma_\beta\left(\frac{\beta+\beta^{-1}}{2} + \frac{\beta}{2}|\epsilon_1 r_1 + \epsilon_2 r_2| + \frac{\beta^{-1}}{2}(\epsilon_1 s_1 + \epsilon_2 s_2) + \epsilon_3 P\right)}.
\end{equation}
For the diagonal insertions $(r_i,s_i)=(0,2\beta P_i)$, it is related to the DOZZ formula by
\begin{equation}
\begin{aligned}
    \widehat{C}^{P}_{(0,2\beta P_1)(0,2\beta P_2)} {}&= 16P \sin(\pi\beta^2)\sin(\pi\beta^{-2}) \Gamma_\beta(2P)\Gamma_\beta(-2P) \\
    & \times \left( \pi\mu\,\frac{\Gamma(b^2)}{\Gamma(1-b^2)}\,b^{2-2b^2} \right)^{(P_1+P_2+P-Q/2)/b} \frac{ \Upsilon_b'(0) \Upsilon_b(2P_1) \Upsilon_b(2P_2) \Upsilon_b(2P)}{C_{\mathrm{DOZZ}}(Q-2P_1,Q-2P_2,Q-2P)}.
\end{aligned}
\end{equation}

\subsection{The conformal block}
For any Young diagram $\nu = (\nu_i)_{i\ge1}$, write
\begin{equation}\label{eq:Ward_coefficient}
    v(h_1,h_2,h_3,\nu) := \prod_{j\ge 1} (\nu_jh_2 - h_1 + h_3 + \sum_{u<j}\nu_u).
\end{equation}
For external weights $\bh=(h_1,h_2,h_3,h_4)$ and internal weight $s$, the sphere four-point conformal block is
\begin{equation}\label{eq:block-definition}
    \mathcal F^{c}_{s}(\bh; z) = z^{s-h_1-h_2} \sum_{n\ge0} B^c_n(s;\bh) z^{n}.
\end{equation}
Here, the coefficients are
\begin{equation}
    B^c_n(s;\bh) = \sum_{|\nu|=|\tilde\nu|=n} v(h_1,h_2,s,\tilde\nu) v(h_4,h_3,s,\nu) G_{s}^{-1}(\nu,\tilde\nu),
\end{equation}
where $G_s$ is the Shapovalov Gram matrix of central charge $c$ and conformal weight $s$, and $G_s^{-1}$ is its inverse.

\subsection{Discrete conformal block expansion for $O(n)$ nesting observables}
The constant $\mathcal{N}_{P_1,P_2}$ in \eqref{eq:O(n)-nesting-formula} is
\begin{equation}
    \mathcal{N}_{P_1,P_2} = \frac{2\beta^{\beta^{-2}-\beta^2} \, \Gamma_\beta(Q)^2 \, \Gamma(\beta^{-2}-1)}{\tan^2(\pi\beta^2) \, \Gamma(1-\beta^2)}
    \prod_{j=1}^{2} \left[ P_j\tan(2\pi\beta P_j) \Gamma_\beta(2P_j)\Gamma_\beta(-2P_j) \right].
\end{equation}

The integrand in \eqref{eq:O(n)-nesting-formula} has three classes of poles: the zeros of the denominator, the poles of the imaginary-DOZZ structure constant $\hat{C}$, and the poles of the conformal block. As explained in \cite{JSW}, the residues contributed from the second and third class of poles cancel pairwise by the Zamolodchikov recursion relation of the block. The remaining poles are $s_m=P_m^2-(1-c)/24$ for $m\in\mathbb{Z}$, with the momentum $P_m$ determined by the loop weight $n_{\mathrm{s}}$ through
\begin{equation}
    n_{\mathrm{s}}=2\cos(2\pi\beta P_{\mathrm{s}}), \qquad P_m=P_{\mathrm{s}}+m\beta^{-1}.
\end{equation}
Evaluating the integral by residues, we have
\begin{equation}
    \mathcal Z^{O(n)}_{h_1,h_2,\ns}(x) = \frac{\mathcal{N}_{P_1,P_2}}{\pi\beta} \cdot \sum_{m\in\mathbb Z} \sin\left(2\pi\beta^{-1}P_m\right) \, \hat{C}^{P_m}_{(0,2\beta P_1)(0,2\beta P_2)} \, \mathcal{F}_{s_m}^c(h_1,h_2,h_2,h_1;x).
\end{equation}
This agrees with \cite[Formula (1.7)]{JSW} and \cite[Formula (2.8)]{DJNRS26}.

\end{document}